\documentclass[useAMS,usenatbib,usegraphicx]{mn2e}

\title[Protogalactic magnetic fields by supernovae]{Strong magnetic fields and large rotation measures in protogalaxies by supernova seeding}
\author[A. M. Beck et al.]{A. M. Beck$^{1,2}$\thanks{E-mail: abeck@usm.uni-muenchen.de}, K. Dolag$^{1,3}$, H. Lesch$^{1}$ and P. P. Kronberg$^{4,5}$ \\
  $^{1}$University Observatory Munich, Scheinerstr. 1, D-81679 Munich, Germany\\
  $^{2}$Max Planck Institute for Extraterrestrial Physics, Giessenbachstr., D-85748 Garching, Germany\\
  $^{3}$Max Planck Institute for Astrophysics, Karl-Schwarzschild-Str. 1, D-85741 Garching, Germany\\
  $^{4}$Theoretical Division, MS B285, Los Alamos National Laboratory, Los Alamos, NM-87545, United States\\
  $^{5}$Department of Physics, University of Toronto, Toronto, Ontario M5S 1A7, Canada}
  
\usepackage[latin1]{inputenc}
\usepackage{float}
\usepackage{graphicx}
\usepackage{amsmath}
\usepackage{amssymb}
\usepackage{amsfonts}
\usepackage{xcolor}
\usepackage{times}

\begin{document}

\date{Accepted 2013 August 15. Received 2013 August 15; in original form 2013 February 25}

\pagerange{\pageref{firstpage}--\pageref{lastpage}} \pubyear{2013}

\maketitle

\label{firstpage}


\begin{abstract}
\noindent{}We present a model for the seeding and evolution of magnetic fields in protogalaxies.
Supernova (SN) explosions during the assembly of a protogalaxy self-consistently provide magnetic seed fields, which are subsequently amplified by compression, shear flows and random motions.
Our model explains the origin of strong magnetic fields of $\mu$G amplitude within the first starforming protogalactic structures shortly after the first stars have formed.

\noindent{}We implement the model into the MHD version of the cosmological N-body / SPH simulation code GADGET and we couple the magnetic seeding directly to the underlying multi-phase description of star formation.
We perform simulations of Milky Way-like galactic halo formation using a standard $\Lambda$CDM cosmology and analyse the strength and distribution of the subsequent evolving magnetic field.

\noindent{}Within starforming regions and given typical dimensions and magnetic field strengths in canonical SN remnants, we inject a dipole-shape magnetic field at a rate of $\approx$10$^{-9}$ G Gyr$^{-1}$.
Subsequently, the magnetic field strength increases exponentially on timescales of a few ten million years within the innermost regions of the halo.
Furthermore, turbulent diffusion, shocks and gas motions transport the magnetic field towards the halo outskirts.
At redshift z$\approx$0, the entire galactic halo is magnetized and the field amplitude is of the order of a few $\mu$G in the center of the halo and $\approx$10$^{-9}$ G at the virial radius.

\noindent{}Additionally, we analyse the intrinsic rotation measure (RM) of the forming galactic halo over redshift.
The mean halo intrinsic RM peaks between redshifts z$\approx$4 and z$\approx$2 and reaches absolute values around $1000$ rad m$^{-2}$.
While the halo virializes towards redshift z$\approx$0, the intrinsic RM values decline to a mean value below $10$ rad m$^{-2}$.
At high redshifts, the distribution of individual starforming and thus magnetized regions is widespread.
This leads to a widespread distribution of large intrinsic RM values.

\noindent{}In our model for the evolution of galactic magnetic fields, the seed magnetic field amplitude and distribution is no longer a free parameter, but determined self-consistently by the star formation process occuring during the formation of cosmic structures.
Thus, it provides a solution to the seed field problem.
\end{abstract}




\section{Introduction}

\noindent{}Radio observations reveal magnetic fields on all scales in the Universe, ranging from small planets to big clusters of galaxies \citep[for reviews on cosmic magnetism see e.g.][]{kronberg94,beck96,widrow02,kulsrud08,vallee11a,vallee11b} and even the largest voids \citep[see e.g.][]{neronov10}.
However, the origin and evolution of the magnetized Universe is still not well understood.
At first, magnetic seed fields must have been created during structure formation.
Afterwards, the seeds were amplified to the observed present-day values and transported to the present-day distribution by a complex interplay of MHD processes.

\noindent{}In the standard cosmological model structures are believed to have assembled in a hierarchical process, with the smallest objects forming first and subsequently merging \citep[on structure formation see e.g. the book of][]{mo10}.
This bottom-up scenario is supported by numerical simulations showing a good agreement between the observed and calculated distribution of large structures.
The process of structure formation can already lead to the creation of magnetic seed fields.

\noindent{}Faraday rotation is a powerful method to measure extragalactic magnetic fields.
It occurs when the plane of polarization of a wave travelling towards the observer is rotated by an intervening magnetic field.
The strength of the effect is described by the rotation measure (RM).
Observations show that galaxies or galactic halos at redshifts z$\gtrsim$2 typically have a widespread distribution of absolute RM values of several $1000$ rad m$^{-2}$.
In contrast, RM values caused by the halo of our Galaxy are around $10$ rad m$^{-2}$ \citep[see e.g.][]{simard81,kronberg82,welter84,carilli94,oren95,carilli97,athreya98,pentericci00,broderick07,kronberg08,mao10,gopal12,hammond12,mao12}.
However, as the RM is the integrated line of sight product of electron density and magnetic field strength, difficulties arise to determine the origin of the large RM.
It is not yet clear, if the observed large RM at high redshifts are caused directly by the sources or by intervening gas clouds with unknown impact parameters along the line of sight.
In any case, the origin and evolution of those RM must be coupled to the formation and evolution of cosmic structures.

\noindent{}Stars are among the earliest objects in the Universe \citep[see e.g.][]{abel02,bromm09}.
Within protogalactic gas clouds battery effects can generate very weak magnetic seed fields.
These small fields are then carried into the newly forming stars, at which point they are enhanced by gravitational compression.
Subsequently, the highly turbulent and fast rotating (proto)stars amplify the seeds by small-scale and $\alpha\omega$-Dynamo action \citep[on dynamos see e.g.][]{shukurov07}.
When the stars explode as supernovae (SN), their magnetic fields are infused together with the gas into the surrounding interstellar medium. 
The strength of the magnetic field within SN remnants is observed to be between $10^{-6}$ and $10^{-3}$ G on scales of order a couple pc \citep{reynolds12}.

\noindent{}Focusing now on galactic scales, where $\mu$G interstellar magnetic fields are common, a variety of processes can be responsible for amplification of the seed fields.
Amplification is assumed to occur mainly by gravitational compression, turbulence and dynamo action.
The most prominent dynamo theory is the mean-field dynamo \citep{krause80}.
However, this dynamo mechanism is challenged by the observations of strong magnetic fields in irregular objects or at very high redshift \citep[see e.g.][]{bernet08,kronberg08}.
Contrarily, fast turbulent dynamos operate on timescales of a few ten million years and lead to an exponential growth of the magnetic energy, first on small scales and then later transported to larger scales \citep[see e.g.][]{zeldovich83,kulsrud92,kulsrud97,mathews97,malyshkin02,brandenburg05,arshakian09,schleicher10}.

\noindent{}Random motions created by the gravitational collapse or injected during structure formation (e.g. by feedback) can drive a small-scale dynamo.
Using analytical calculations and numerical simulations, \cite{beck12} showed that the process of galactic halo formation and virialization \citep[see also][]{wise07} is sufficient to enhance primordial magnetic fields up to the observed $\mu$G values.
In their model, magnetic perturbations are amplified by turbulent motions until the point where equipartition is reached between the magnetic and turbulent energy density.
Additionally, \cite{kotarba11} and \cite{geng12a,geng12b} show the amplification of magnetic fields in major and minor galactic mergers.
Both accompany structure formation, especially at high redshifts \citep{somerville00}.
Recently, the first simulations of galaxy formation have been carried out including the evolution of magnetic fields \citep[see e.g.][]{wang09,beck12,pakmor12,latif13}.
However, these simulations assumed the magnetic field to be of a primordial origin and did not seed them within the simulations.

\begin{figure}
\begin{center}
  \includegraphics[bb= 302 497 628 932, width=0.45\textwidth]{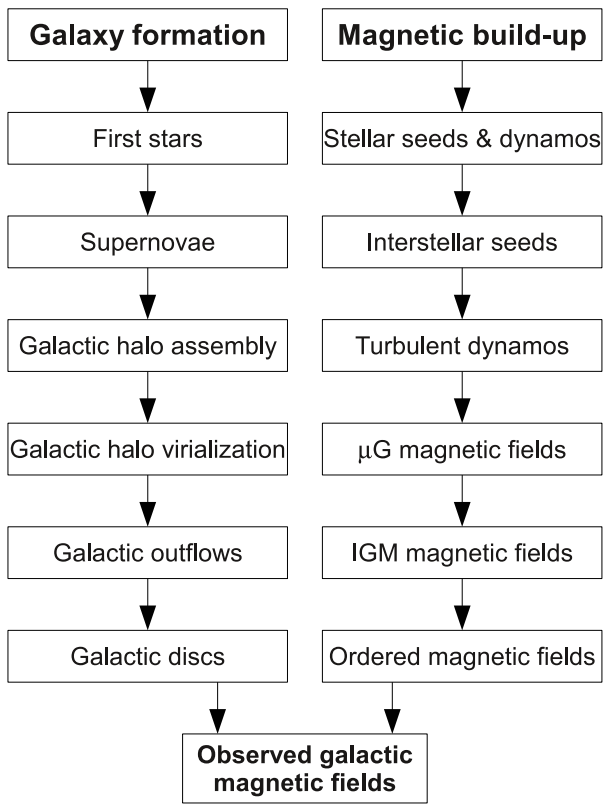}
  \caption{Schematic overview of the process of SN magnetic seeding, amplification and ordering during structure formation.
Note that various other seeding mechanisms can mutually co-operate and provide local (e.g. AGN) or global (e.g. primordial) magnetic seed fields.}
  \label{fig:structure}
\end{center}
\end{figure}

\noindent{}A more consistent description of the origin of galactic magnetic fields needs to incorporate SN-created seed fields.
Their existence is independently verified and the resulting amplification, diffusion and gas motions have been calculated and discussed in several papers \citep[see e.g.][]{bisnovatyi73,rees87,pudritz89,kronberg99,rees94,rees06,chyzy11}.
A schematic overview of this magnetic build-up scenario during galaxy formation is shown in Fig. \ref{fig:structure}.

\noindent{}In principle, a similar scenario can  be constructed with active galactic nuclei (AGN).
Within the highly conducting accretion discs surrounding supermassive black holes, magnetic fields can be easily seeded by battery processes and amplified on very short dynamical timescales.
As indicated by observations of radio galaxies, the magnetized material is transported into the intergalactic medium (IGM) along powerful jets.
This magnetized material can then mix with the galactic gas content and provide a magnetic seed field within the galaxies \citep[see e.g.][]{willis78,strom80,kronberg94,furlanetto01,kronberg01,rees06,kronberg09,colgate11}.
The magnetic energy provided by an AGN can, if compressed into the volume of a galaxy, lead to $\mu$G magnetic field amplitudes \citep[see][]{daly90,kronberg01}.
So far, simulations with AGN seeding and subsequent evolution have been mostly applied to the magnetization of the IGM of galaxy clusters \citep[see e.g.][]{xu08,xu10,xu12}.
However, within galaxies, the first generation of stars can provide magnetic seed fields earlier than the first generation of AGN.

\noindent{}In this work, we present a numerical model for the seeding of magnetic fields by SN explosions.
We repeat previous cosmological simulations of Milky Way-like galactic halo formation \citep{beck12}, by incorporating our new SN seeding model.
We analyse the distribution and amplitude of the resulting halo magnetic field.
We also study the redshift evolution of the intrinsic RM of the galactic halo.

\noindent{}The paper is organized as follows.
The numerical method and initial conditions are briefly described in section 2.
Section 3 introduces the new seeding model.
An analysis of our simulations, including the magnetic field evolution and the resulting intrinsic RM is given in section 4.
The main results are summarized in section 5.


\section{Numerics}

\subsection{Numerical method}

\noindent{}We apply the same numerical method as already used in \cite{beck12}.
The simulations are performed with the N-body / SPMHD code GADGET \citep{springel01a,springel05a,dolag09b}.
This code uses a formulation of SPH which conserves both energy and entropy \citep{springel02}.
For a recent review on the SPH and SPMHD method see \cite{price12}.
On-the-fly calculation of center positions and virial radii of bound structures (haloes and subhaloes) is done with SUBFIND \citep{springel01b,dolag09a}.

\begin{table}
\begin{center}
  \begin{tabular}{@{}lll}
    \hline\hline
    &\hspace*{-2.8cm}\textsc{Multi-phase Model Parameters}&\\\hline\hline
    Gas consumption timescale & t$_{\rmn{SF}}$ & 2.1 Gyr\\
    Gas density threshold & n$_{\rmn{th}}$ & $0.13$ cm$^{-3}$\\
    SN mass fraction & $\beta$ & $0.1$\\
    SN per solarmass formed & $\alpha$ & $0.008$ $M_{\odot}^{-1}$\\
    Evaporation parameter & A & 1000\\
    Effective SN temperature & T$_{\rmn{SN}}$ & $10^{8}$ K\\
    Temperature of cold clouds & T$_{\rmn{CC}}$ & 1000 K\\\hline\hline
  \end{tabular}
  \caption{Parameters of the star formation model \citep{springel03} used in the simulations.}
  \label{tab:sfr_model}
\end{center}
\end{table}

\noindent{}Furthermore, we apply the \cite{springel03} star formation model without the implementation of galactic winds.
It describes radiative cooling, UV background heating and SN feedback in a consistent two-phase sub-resolution model for the interstellar medium.
Cold clouds with a fixed temperature of $T_{\rmn{CC}}$ are embedded into a hot ambient medium at pressure equilibrium.
These cold clouds will evaporate with an efficiency parameter of $A$ and form stars on a timescale of $t_{\rmn{SF}}$ at a density threshold of $\rho_{\rmn{th}}$.
A fraction $\beta$ of these stars is expected to die instantly as SN, thereby heating the gas with a temperature component of $T_{\rmn{SN}}$.
Additionally, the gas is loosing energy via radiative cooling, modeled by assuming a primordial gas composition and a temperature floor of $50$ K.
The cooling depends only on density and temperature and not on metallicity.
This leads to a self-regulated cycle of cooling, star formation and feedback within the gas.
The numerical values for the star formation model in our simulations are given in Table 1.

\noindent{}Ideal MHD is implemented into GADGET following \cite{dolag09b} using the standard (direct) method, where the magnetic field is evolved by the induction equation and reacts back on the gas with the Lorentz force.
Additionally, we need to model the diffusion of the magnetic field from the starforming regions into the surrounding medium in a physically plausible way.
Our approach is based on the implementation of magnetic resistivity in GADGET as described in \cite{bonafede11} using the diffusion coefficient $\eta$ at an assumed constant value of $10^{27}$cm$^{2}$ s$^{-1}$.
This value is reasonable for galactic scales \citep{longair10}.
Expect for the magnetic resistivity, we do not use an explict scheme to control the magnetic divergence (such as the Dedner cleaning \citep[][]{dedner02,stasyszyn13}, or any Euler potential method).
The resistivity is represented by an effective sub-grid model for the turbulent magnetic field decay and its value is larger than those of numerical or ohmic resistivity.
However, as in our case the initial magnetic fields are very localized within starforming regions and we need to carefully handle cases where magnetic fields are diffused outside the starforming regions.
This can happen in an implausible and unphysical way.
Hence, we limit the distance over which the magnetic field is transported in each timestep, depending on the local diffusion velocity, diffusion length and timestep.
Within our cosmological simulations we employ the transport of the magnetic field at each timestep in the following way:

\noindent{}Each SPMHD particle has an unique timestep $\Delta{}t$ and smoothing length $h$ (i.e. resolution length).
First, we estimate the local diffusion speed 
\begin{equation}V_\rmn{D}=\sqrt{\frac{1}{2}\left(c_\rmn{s}^{2}+v_\rmn{a}^{2}\right)}\end{equation}
as the square root mean of the local sound- and Alfv\'{e}n speed.
Secondly, we estimate the distance $L_\rmn{D}$ over which diffusion is taking place 
\begin{equation}L_\rmn{D}=V_\rmn{D}\Delta{}t\end{equation}
as the product of local diffusion speed and timestep.
If the local diffusion distance $L_\rmn{D}$ is larger than our spatial resolution element (given by the smoothing length $h$), the normal diffusion coefficient $\eta$ is used.
If $L_\rmn{D}$ is smaller than our spatial resolution element, we follow a stochastic approach in modelling the magnetic diffusion.
In analogy to the stochastic star formation algorithm, we draw a random number from the interval $[0,1[$ and compare it to the quotient $L_\rmn{D}/h$ of diffusion distance and smoothing length.
If the random number exceeds the quotient $L_\rmn{D}/h$ normal diffusion is performed.
Otherwise, we switch off the diffusion by setting the diffusion coefficient $\eta$ to zero for this timestep.
Furthermore, we employ a minimum value of 5 km s$^{-1}$ for the diffusion velocity.
This new diffusion model allows us to mimic the transport of magnetic energy outside the starforming regions in a conservative way and successfully suppresses numerical diffusion of the magnetic field into low density regions.

\subsection{Initial conditions}

\noindent{}We use the same initial conditions as in \cite{beck12}, which are originally introduced in \cite{stoehr02}.
Out of a large cosmological box, with a power spectrum index for the initial fluctuations of $n=1$ and an amplitude of $\sigma_{8}=0.9$ a Milky Way-like dark matter halo is identified.
Simulations with different resolutions of this halo are created and labelled GA0, GA1 and GA2, containing 13~603, 123~775 and 1~055~083 dark matter particles inside the virial radius.
The forming dark matter halo is comparable in mass to the halo of the Milky Way ($\approx3\times10^{12} M_{\odot}$) and in virial size ($\approx 270$ kpc).
The halo does not undergo any major merger after a redshift of z$\approx$2 and also hosts a subhalo population comparable to the satellite population of the Milky Way.
In order to add a baryonic component, the high resolution dark matter particles are split into an equal amount of gas and DM particles.
The mass of the initial DM particle is split according to the cosmic baryon fraction, conserving the center of mass and the momentum of the parent DM particle.
The new particles are displaced by half the mean interparticle distance.


\section{Magnetic Seeding model}

\noindent{}In this section we present the SN seeding model.
We describe the numerical model for the amplitude of the magnetic energy injection, as well as the corresponding dipole structure.
\noindent{}Starting with the induction equation of non-ideal MHD we include a time-dependent seeding term on the right hand side of the induction equation in addition to the convective and (spatially constant) resistive term, resulting in

\begin{equation}\frac{\partial{}\bmath{B}}{\partial{}t}=\bmath{\nabla}\times\left(\bmath{v}\times\bmath{B}\right)+\eta\Delta\bmath{B}+\left.\frac{\partial{}\bmath{B}}{\partial{}t}\right|_{\rmn{Seed}}.\label{equ:induction}\end{equation}

\noindent{}The magnetic seeding amplitude per timestep $\Delta{}t$ is given by

\begin{equation}\left.\frac{\partial{}\bmath{B}}{\partial{}t}\right|_{\rmn{Seed}}=\sqrt{N^\rmn{eff}_\rmn{SN}}\frac{B_{\rmn{Inj}}}{\Delta{t}}\bmath{e}_\rmn{B},\end{equation}

\noindent{}where $\bmath{e}_\rmn{B}$ is a unity vector and $B_{\rmn{Inj}}$ defines the injected magnetic field amplitude and $N^\rmn{eff}_\rmn{SN}$ is a normalisation constant, which specifies the effective number of SN explosions.
The number of SN explosions is not a free input parameter into our model but it is calculated directly from the sub-grid model for star formation.
Our simulations include star formation via the \cite{springel03} model, however, we note that our seeding model can easily be coupled to any other star formation model.
The mass of cold gas turning into stars, per timestep, is

\begin{equation}m_\rmn{\ast}=\frac{\Delta{}t}{t_{\ast}}m_\rmn{c},\end{equation}

\noindent{}with $m_\rmn{c}$ the mass of cold gas available for star formation.
Within the adapted star formation model, a total gas consumption timescale $t_{\rmn{SF}}$ is scaled by the gas density and a density threshold to yield the local star formation timescale

\begin{equation}t_{\ast}=t_{\rmn{SF}}\left(\frac{\rho_\rmn{th}}{\rho}\right)^{\frac{1}{2}}.\end{equation}

\noindent{}The effective number of SN explosions is given by:

\begin{equation}N^\rmn{eff}_\rmn{SN}=\alpha{}m_\rmn{\ast},\end{equation}

\noindent{}where the number of SN events per formed solar mass in stars is specified by $\alpha$. Its numerical value can be calculated from the initial mass function \citep[see][]{hernquist03} and the corresponding value can be found in Table \ref{tab:sfr_model}.
We calculate the total injected magnetic field amplitude for all SN events to be

\begin{equation}B_\rmn{Inj}^\rmn{all}=\sqrt{N^\rmn{eff}_\rmn{SN}}B_\rmn{SN}\left(\frac{r_\rmn{SN}}{r_\rmn{SB}}\right)^{2}\left(\frac{r_\rmn{SB}}{r_\rmn{Inj}}\right)^{3}.\end{equation}

\noindent{}Here, $B_\rmn{SN}$ is the mean magnetic field strength within one SN remnant of radius $r_\rmn{SN}$.
We assume a spherical geometry for the remnant and isotropically expand the remnants magnetic field from the initial radius into a bubble with radius $r_\rmn{SB}$.
The bubbles are then randomly placed and mixed within a sphere of radius $r_\rmn{Inj}$ \citep[for a similar scaling see][]{hogan83}.
The size of the injection sphere is determined by the size of the numerical resolution elements (i.e. the SPMHD the smoothing length).

\noindent{}The magnetic field seeding rate that results from this model can be extrapolated from

\begin{equation}\dot{B}_\rmn{seed}\approx{}B_\rmn{SN}\left(\frac{r_\rmn{SN}}{r_\rmn{SB}}\right)^{2}\left(\frac{r_\rmn{SB}}{r_\rmn{Inj}}\right)^{3}\frac{\sqrt{\dot{N}_\rmn{SN}\Delta{}t}}{\Delta{t}},\label{equ:approx}\end{equation}

\noindent{}where $\dot{N}_\rmn{SN}$ is the SN occurance rate. 
From \cite{reynolds12} we take for a canonical SN remnant a radius of $r_\rmn{SN}=5$ pc and a mean field strength of $B_\rmn{SN}=10^{-4}$ G, which we assume is afterwards blown into bubbles of $r_\rmn{SB}=25$ pc.
Using Eq. \ref{equ:approx} we can derive a quick estimate of the mean seeding rate of the Milky Way (volume rougly 300 kpc$^3$) during its lifetime of presumably 10 Gyr.
If about $10^8$ SN occured within our Galaxy, we find a magnetic seeding rate of $\approx{}10^{-26}$ G s$^{-1}$ or $\approx{}10^{-9}$ G Gyr$^{-1}$ \citep[i.e. see also][]{rees94}.
This estimate only takes into account the mixing of the past SN seed events, however, it neglects the subsequent evolution of the magnetic field.
This can lead to amplification, distribution or dilution.

\noindent{}The magnetic field injected by the seeding term must be divergence-free.
Hence, it is natural to choose a dipole structure with a dipole moment $\bmath{m}$ and then the seeding term takes the form:

\begin{equation}\left.\frac{\partial{}\bmath{B}}{\partial{}t}\right|_\rmn{Seed}=\frac{1}{|\bmath{r}|^{3}}\left(3\left(\frac{\partial{}\bmath{m}}{\partial{}t}\cdot\bmath{e}_{r}\right)\bmath{e}_{r}-\frac{\partial{}\bmath{m}}{\partial{}t}\right),\end{equation}

\noindent{}with $\bmath{e}_{r}$ the unity vector in $\bmath{r}$ direction.
Each starforming region will create a magnetic dipole around itself.
The time derivative of each dipole moment is then given by

\begin{equation}\frac{\partial{}\bmath{m}}{\partial{}t}=\sigma\frac{B^\rmn{all}_{\rmn{Inj}}}{\Delta{t}}\bmath{e}_\rmn{B},\end{equation}

\noindent{}where $\sigma$ is a numerical normalisation constant and $\bmath{e}_\rmn{B}=\bmath{a}/|\bmath{a}|$ a unit vector defining the direction of the dipole moment.
In the weak-field approximation we can choose the direction of the acceleration field $\bmath{a}$.

\noindent{}In our numerical model, the magnetic dipoles do not extend to infinity.
They are softened at the center and truncated at the injection scale and the energy of the numerical dipoles has to be renormalised.
Softening of the dipole is necessary in order to avoid discontinuities for $|\bmath{r}|\rightarrow{}0$.
By integration over the modified volume we follow \cite{donnert09} and find for the normalisation constant

\begin{equation}\sigma=r_\rmn{Inj}^{3}\sqrt{\frac{1}{2}f^3(1+f^{3})},\end{equation}

\noindent{}where $f=r_\rmn{soft}/r_\rmn{Inj}$ is the ratio between dipole softening length and truncation length (i.e. in SPMHD the smoothing length).
In our implementation in the code, the truncation length is also the injection length $r_\rmn{Inj}$.
During the simulations, a magnetic field is injected onto its neighbour particles in a dipole shape for each starforming gas particle.
Overlapping dipoles are added linearly.


\section{Simulations}

\noindent{}This section presents the results of our cosmological simulations.
Contour images of the different physical quantities and the calculation of intrinsic RM values are created by projecting the SPMHD data in a comoving $(1$ $\rmn{Mpc})^{3}$ cube, centered on the largest progenitor halo onto a $512^{2}$ grid using the code P-SMAC2 (Donnert et al., in preparation).
The precise details of the projection algorithm can be found in \cite{dolag05}.
In principle, we calculate the overlap of each particle with each line of sight and integrate:

\begin{equation}A_\rmn{proj}=\sigma{}\int{\left[\sum{\frac{m_j}{\rho_j}A_{j}W\left[d_{j}(r)/h_{j}\right]}\right]dr},\end{equation}

\noindent{}where $A$ is the quantity of interest, $\sigma$ the integral normalisation, $m$ and $\rho$ the particle mass and density, $W$ and $h$ the SPMHD kernel function and smoothing length and $d(r)$ the element of distance with respect to the position $r$ along the line of sight.
Table \ref{tab:setup_sims} shows the performed simulations:
Initial conditions of different resolution (GA0, GA1 and GA2) are used to study the seeding model and subsequent amplification.

\begin{table*}
\begin{center}
  \begin{tabular}{@{}lllllllll}
    \hline\hline
    &&&&\hspace*{-1cm}\textsc{Simulation setup}&\\
    \hline\hline
    Scenario & $\rmn{N}_{\rmn{Gas}}$ & $\rmn{N}_{\rmn{DM}}$ & $\rmn{M}_{\rmn{Gas}}$ & $\rmn{M}_{\rmn{DM}}$ & $\rmn{B}_{\rmn{Strength}}^{\rmn{SN}}$ & $\rmn{R}_{\rmn{Radius}}^{\rmn{SN}}$ & $\rmn{R}_{\rmn{Radius}}^{\rmn{Bubble}}$ & $\rmn{f}_{\rmn{Soft}}^{\rmn{SN}}$\\
    \hline
    ga0\_seed\_all & 68323 & 68323 & $2.6\cdot{}10^{7}$ $\rmn{M}_{\odot}$ & $1.4\cdot{}10^{8}$ $\rmn{M}_{\odot}$ & $10^{-4}$ G & $5$ pc & $25$ pc &$0.25\cdot{}h$\\
    ga1\_seed\_all & 637966 & 637966 & $2.8\cdot{}10^{6}$ $\rmn{M}_{\odot}$ & $1.5\cdot{}10^{7}$ $\rmn{M}_{\odot}$ & $10^{-4}$ G & $5$ pc & $25$ pc &$0.25\cdot{}h$\\
    ga2\_seed\_all & 5953033 & 5953033 & $3.0\cdot{}10^{5}$ $\rmn{M}_{\odot}$ & $1.6\cdot{}10^{6}$ $\rmn{M}_{\odot}$ & $10^{-4}$ G & $5$ pc & $25$ pc &$0.25\cdot{}h$\\
    \hline
    ga2\_seed\_low & 5953033 & 5953033 & $3.0\cdot{}10^{5}$ $\rmn{M}_{\odot}$ & $1.6\cdot{}10^{6}$ $\rmn{M}_{\odot}$ & $10^{-5}$ G & $5$ pc & $25$ pc &$0.25\cdot{}h$\\
    ga2\_seed\_high & 5953033 & 5953033 & $3.0\cdot{}10^{5}$ $\rmn{M}_{\odot}$ & $1.6\cdot{}10^{6}$ $\rmn{M}_{\odot}$ & $10^{-3}$ G & $5$ pc & $25$ pc &$0.25\cdot{}h$\\
    ga2\_primordial & 5953033 & 5953033 & $3.0\cdot{}10^{5}$ $\rmn{M}_{\odot}$ & $1.6\cdot{}10^{6}$ $\rmn{M}_{\odot}$ & \multicolumn{4}{|c|}{$\rmn{B}_\rmn{primordial}=10^{-10}$ G}\\
    \hline\hline
  \end{tabular}
  \caption{Setup of the different simulations: The table lists the number of gas and dark matter particles, the mass of the gas and dark matter particles, the initial SN remnant radius and magnetic seed field strength, as well as the numerical softening length (where $h$ is the SPMHD smoothing length).}
  \label{tab:setup_sims}
\end{center}
\end{table*}

\subsection{Morphological evolution}

\begin{figure*}
\begin{center}
  \includegraphics[width=0.95\textwidth]{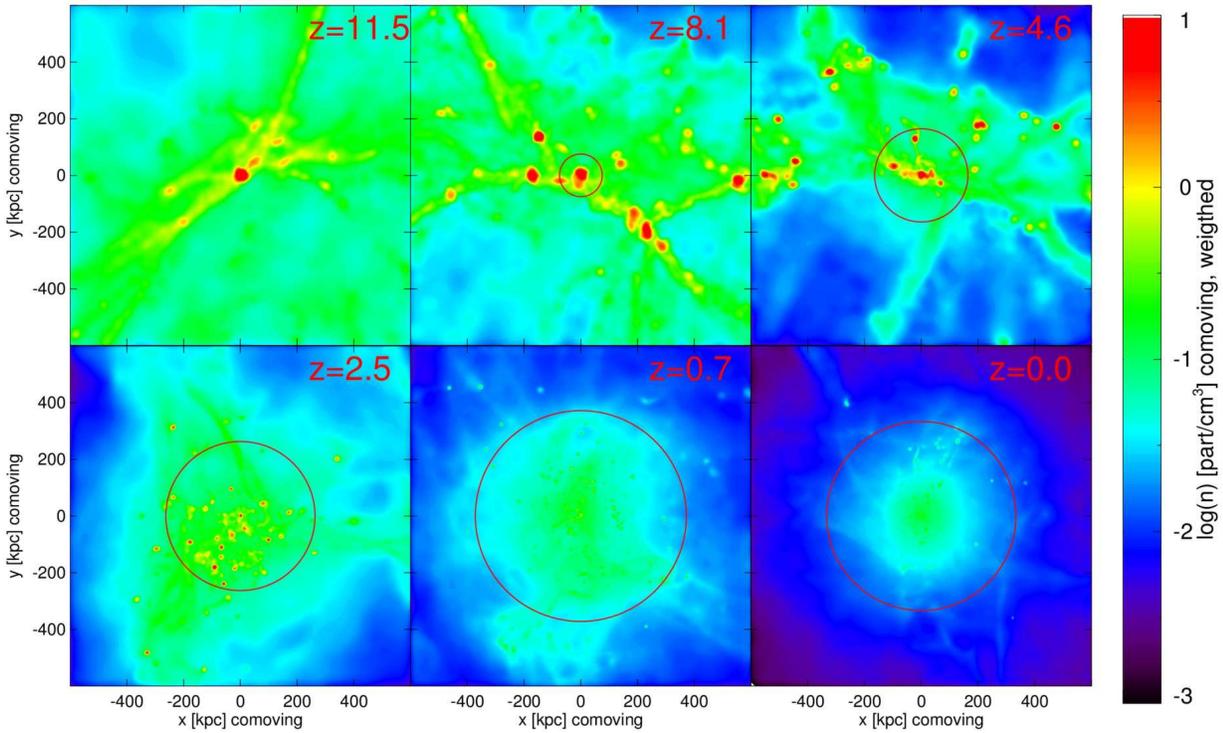}
  \caption{Projected and weighted (integrated over 1 Mpc) gas density $n_{\rmn{gas}}$ in comoving units at different redshifts in the simulation ga2\_seed\_all centered on the halo center of mass.
The red circles indicate the virial radius of the halo.
The formation of filaments and protohaloes with subsequent merger events is visible.}
  \label{fig:sim_dens}
\end{center}
\end{figure*}

\begin{figure*}
\begin{center}
  \includegraphics[width=0.95\textwidth]{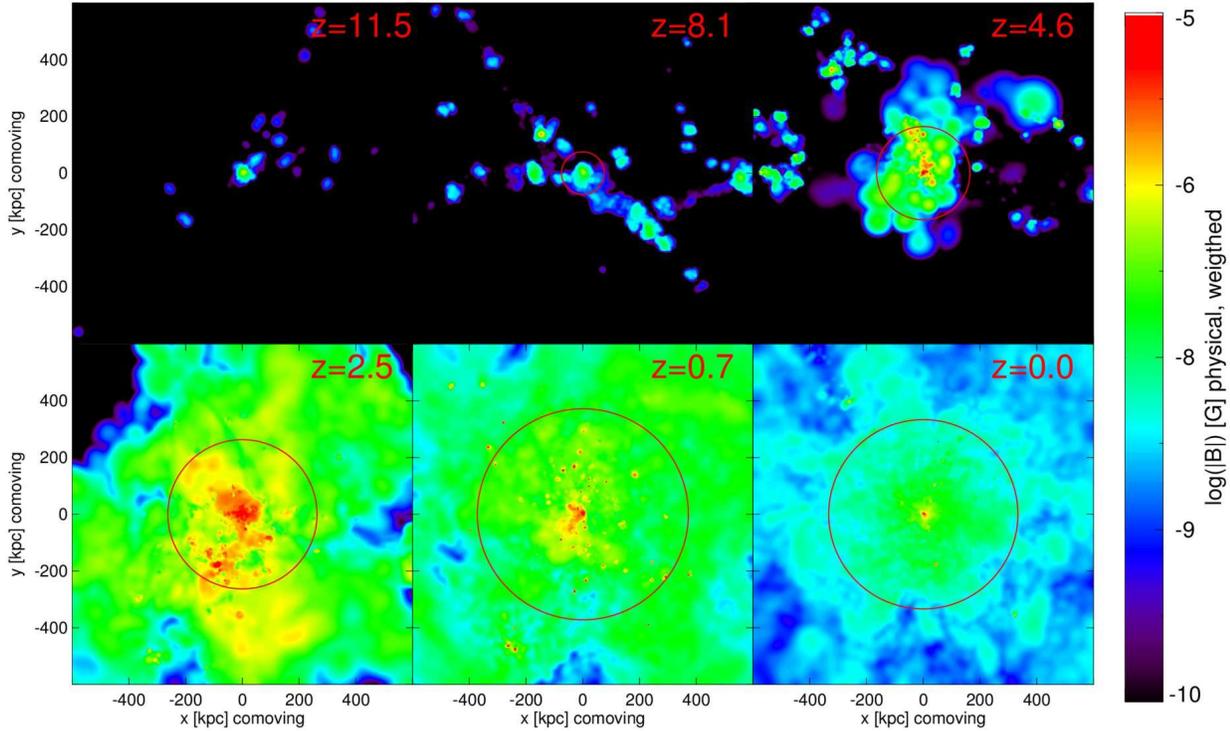}
  \caption{Projected and weighted (integrated over 1 Mpc) total magnetic field strength in physical units at different redshifts in the simulation ga2\_seed\_all centered on the halo center of mass.
The red circles indicate the virial radius of the halo.
Seeding and amplification of the magnetic field within starforming protohaloes is visible.
Furthermore, gas motions and diffusion are carrying magnetic energy towards the halo outskirts.}
  \label{fig:sim_bfld}
\end{center}
\end{figure*}

\begin{figure}
\begin{center}
  \includegraphics[bb= 170 315 769 861, width=0.45\textwidth]{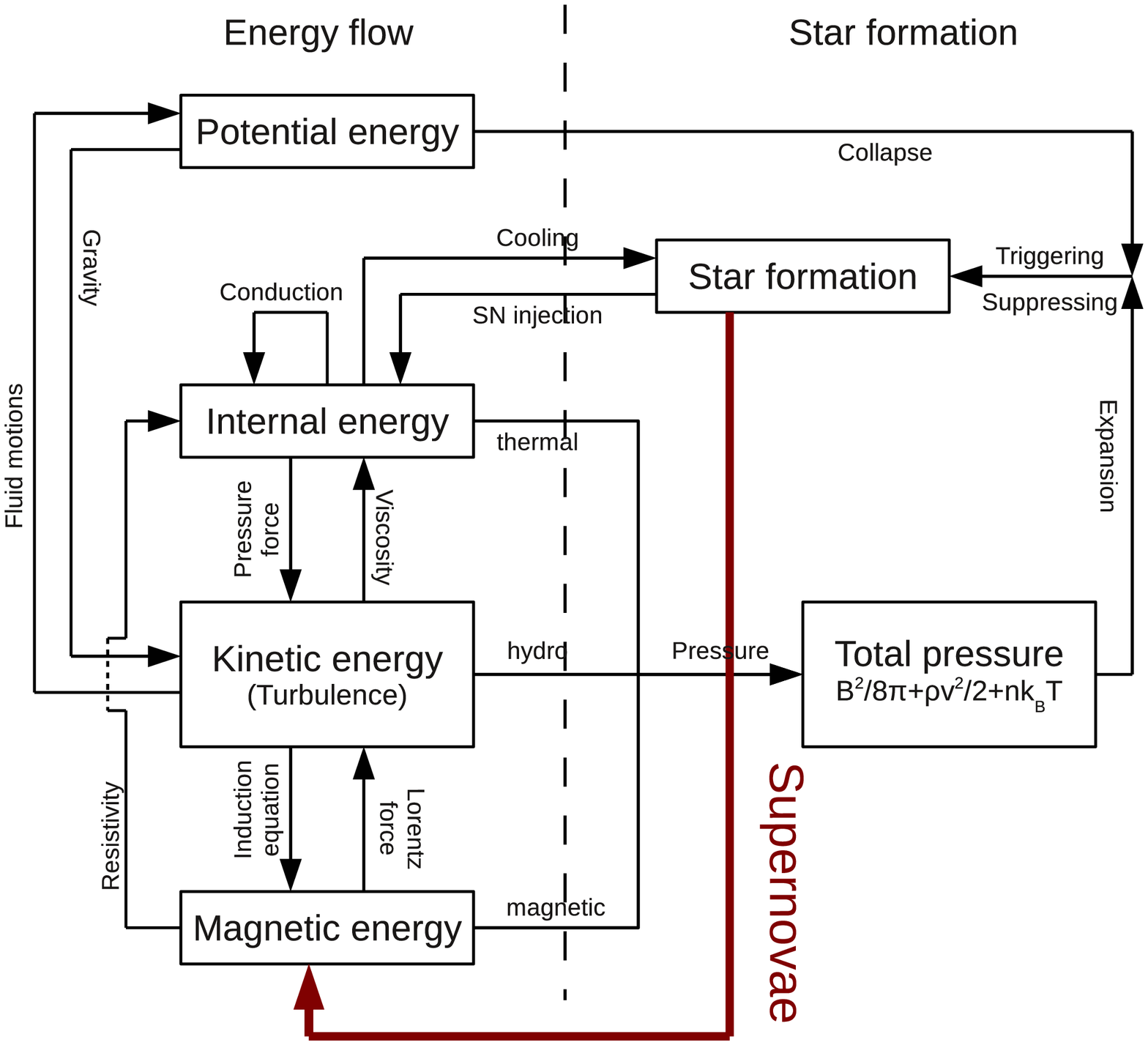}
  \caption{Flows of energy within our simulations.
The diagram is an adaption of Fig. 7 of \citet{beck12} and shows the additional connection between star formation and magnetic energy given by our new SN seeding model.}
  \label{fig:energy_flow}
\end{center}
\end{figure}

\begin{figure}
\begin{center}
  \includegraphics[width=0.45\textwidth]{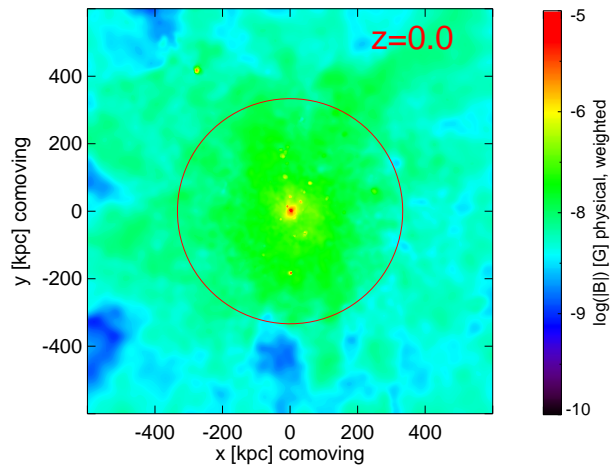}
  \caption{Projected and weighted (integrated over 1 Mpc) total magnetic field strength in physical units at redshift z$\approx$0 in the simulation ga2\_primordial.
The plot corresponds to the bottom right panel of Fig. \ref{fig:sim_bfld}, just with a primordial seed field of $B_\rmn{primordial}=10^{-10}$ G and no SN seeding is applied.}
  \label{fig:sim_prim}
\end{center}
\end{figure}

\noindent{}Fig. \ref{fig:sim_dens} shows the projected and weighted gas density of the gas at six different redshifts in the simulation ga2\_seed\_all, together with the corresponding virial radii of the forming galactic halo.
Fig. \ref{fig:sim_bfld} shows the corresponding projected and weighted total magnetic field strength.
The different phases during formation of the halo and the magnetic seeding, amplification and transport can clearly be seen.
Within the first protohaloes and filaments we find star formation to set in at a redshift z$\approx$20 and at that time the first magnetic seed fields are also created.
The magnetic field is seeded at an amplitude of $\approx{}10^{-9}$ G, consistent with our expectations from section 3.
In the regions without star formation no magnetic field is yet present at these high redshifts.
As long as stars continue to form magnetic fields are seeded within the simulation, however, given the decrease of the star formation rate towards redshift z$\approx$0, correspondingly less magnetic energy is injected.
Subsequently, the seed fields are amplified by gravitational compression and turbulent dynamo action within the first structures \citep[for the amplification mechanism see also][]{beck12}.
This leads to $\mu$G magnetic fields within the first protogalactic objects shortly after they form.
Once seeded, the magnetic field is subject to amplification and diffusion and the contributions of additional SN can be neglected.
Afterwards, the magnetic field starts to diffuse outwards from the starforming regions and towards outer regions of the halo, thereby enriching the IGM with magnetic seed fields.
Furthermore, while the assembly of the main galactic halo continues and merger events are taking place, shockwaves are propagating outwards from the halo center.
The magnetic fields within the IGM are subsequently amplified by merger-driven shockwaves propagating into the IGM.
Within each shockwave, the magnetic field can be amplified by compression at the shockfront and possible dynamo action behind the shockfront \citep[see e.g.][]{kotarba11}.
Around a redshift of z$\approx$2, the last major merger event takes place and the magnetic field saturates, i.e. it evolves into energy density equipartition.
The magnetic field saturates within the innermost regions at a value of a few $\mu$G and at the halo outskirts at about $\approx{}10^{-9}$ G, both at redshift z$\approx$0.

\noindent{}Fig. \ref{fig:sim_prim} shows the result of a simulation run with a primordial seed field of $B_\rmn{prim}=10^{-10}$ G but the new diffusion model.
The amplitude and distribution of the magnetic field at redshift z$\approx$0 obtained by evolving a primordial seed field or by SN seeding and subsequent evolution are almost indistinguishable within the halo.
Outside the halo, the amplitude is slightly higher in the simulation with a primordial seed field than in the simulation with the SN seeding model.
For a detailed study of primordial seed fields during galactic halo formation we refer to our old simulations described in \cite{beck12}.
Fig. \ref{fig:energy_flow} shows how the simulations have been modified.
The most significant differences are:
In the simulations of \cite{beck12} a primordial seed field is already present everywhere at high redshifts and the field is subject to amplification by the very first occurences of compression, random motions and shockwaves.
By contrast, in our new simulations with SN seeding (this work), the magnetic seed field has to be created first within the starforming regions, before it can be subsequently amplified.
Starformation is a new source of magnetic energy in addition to its property as a source and sink of thermal energy.
Furthermore, to create an IGM magnetic field, diffusion and gas motions have to transport the magnetic field into the IGM first, before subsequent amplification can take place.
Our new simulations do still not form a galactic disk, however, we note that the prior presence of $\mu$G magnetic fields at the center of the halo is sufficient to demonstrate that a disk at the center could host magnetic fields of similar strength.

\subsection{Magnetic amplitude and filling factors}

\begin{figure}
\begin{center}
  \includegraphics[angle=90,width=0.475\textwidth]{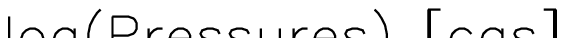}
  \caption{Volume-weighted energy densities as a function of redshift in the simulation ga2\_seed\_all inside the largest progenitor halo.
The magnetic energy density gets seeded and amplified during the phase of galactic halo formation until it reaches equipartition with the corresponding energy densities, particularly the turbulent energy density.}
  \label{fig:sim_equi}
\end{center}
\end{figure}

\begin{figure}
\begin{center}
  \includegraphics[angle=90,width=0.475\textwidth]{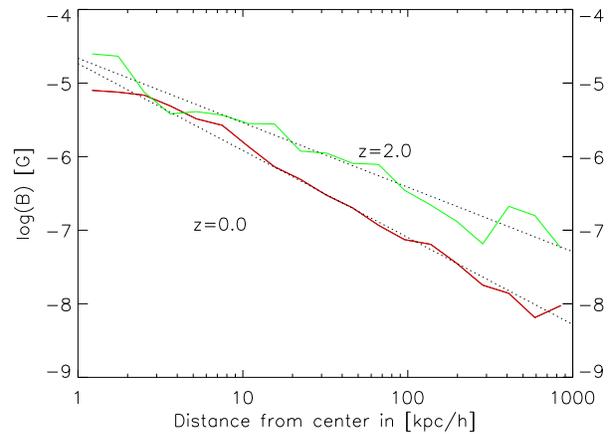}
  \caption{Radial profiles of the mass-weighted magnetic field strength inside the largest progenitor halo in the simulation ga2\_seed\_all for redshifts z$\approx$2 and z$\approx$0, respectively.
The slope of the magnetic field at redshift z$\approx$2 is about $-0.9$ and about $-1.2$ at redshift z$\approx$0.}
  \label{fig:sim_prof}
\end{center}
\end{figure}

\begin{figure}
\begin{center}
  \includegraphics[angle=90,width=0.475\textwidth]{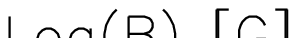}
  \caption{Volume-weighted magnetic field strength as a function of redshift in different types of simulations inside the largest progenitor halo.
The results from the primordial seed field and from the SN seeded fields agree well.
The seeding by SN alone seems not sufficient for generating strong magnetic fields and additional amplification or injection via AGN \citep[see e.g.][]{daly90,kronberg01} is necessary.}
  \label{fig:sim_amp}
\end{center}
\end{figure}

\noindent{}Fig. \ref{fig:sim_equi} shows the magnetic energy density $\varepsilon_{\rmn{mag}}=B^{2}/8\pi$, the kinetic energy density $\varepsilon_{\rmn{kin}}=\rho{}v^{2}/2$, the thermal energy density $\varepsilon_{\rmn{therm}}=(\gamma-1)\rho{}u$ and the turbulent energy density $\varepsilon_{\rmn{turb}}=\rho{}v_{\rmn{turb}}^{2}/2$ in the simulation ga2\_seed\_all.
The adiabatic index $\gamma$ is $5/3$ and $u$ denotes the internal energy.
We estimate the turbulent velocity $v_{\rmn{turb}}$ similar to \cite{kotarba11} and \cite{beck12}.
It is locally approximated by calculating the RMS velocity within the smoothing sphere of each individual SPMHD particle.
\cite{kotarba11} claim it to be a good approximation of the turbulent velocity, although it might slightly overestimate the turbulence on small scales and it ignores turbulence on scales larger than the smoothing scale.
We volume-weight the energy densities among the particles according to

\begin{equation}\bar{\epsilon}=\sum_{j}{\left(\epsilon_{j}\frac{m_j}{\rho_j}\right)} / \sum_{j}{\left(\frac{m_j}{\rho_j}\right)},\end{equation}

\noindent{}where $m$ and $\rho$ are the local mass and gas density and represent the particle volume.
At the beginning of the simulation the magnetic energy density is seeded at values significantly lower than the other energy densities.
However, as the halo begins to assemble the magnetic energy density is amplified to an equipartition value by compression and random motions created by the gravitational collapse.
The magnetic energy density remains then within the same order of magnitude as the thermal and turbulent energy densities.
After the last merger event the entire halo virializes and irregularities within the velocity and magnetic fields are dissipated, leading to decaying energy densities.

\noindent{}Fig. \ref{fig:sim_prof} shows radial profiles of the mass-weighted magnetic field strength inside the galactic halo for two different redshifts.
At both times, magnetic field strengths of $\approx{}10\mu$G are reached in the innermost parts of the halo, however, at redshift z$\approx$0 the magnetic field strength at the halo outskirts is slightly lower.
Outside the halo the magnetic field decreases, because forces maintaining the magnetic field efficiently operate only at the halo center.
Furthermore, after the halo has assembled, the effect of shockwaves contributing to the amplification of the IGM magnetic field is significantly reduced.

\noindent{}Fig. \ref{fig:sim_amp} shows the magnetic field strength within the halo for different types of simulations.
In the simulation with the primordial seed field (starting value of $10^{-10}$ G) the magnetic field gets amplified during the formation of the halo until it saturates around a redshift of z$\approx$2 at values of a few $\mu$G.
Subsequently, the magnetic field is subject to diffusion, which leads to a decay of the amplitude until redshift z$\approx$0 to values of a few $10^{-8}$ G.
In the simulations performed with SN seeding, the magnetic field is first seeded at an amplitude of $\approx{}10^{-9}$ G Gyr$^{-1}$ and then behaves qualitatively similar to the run with primordial seeding.
However, in the simulation with the primordial seed field, the amplification is more efficient and the saturation values at high and at low redshifts are slightly higher.
In simulations with higher resolution, the magnetic field is amplified more efficiently and it saturates at a slightly higher amplitude.

\noindent{}Fig. \ref{fig:sim_amp} also shows a fictive growth curve corresponding to the case where the only source of magnetic energy is SN seeding.
We model this case by calculating the cumulative seeding rate over time as approximated by Eq. \ref{equ:approx} with the SN rate normalized to the Milky Way and evolving with time as suggested by \cite{hernquist03}.
Furthermore, we include cosmological dilution as described in \cite{beck12} and turbulent diffusion on a timescale of the Hubble time, but we neglect possible post-amplification processes \citep[see e.g.][]{ryu08}.
We conclude that the seeding by SN only is not able to build up strong magnetic fields at high redshifts and also leads to a significantly lower amplitude at z$\approx$0.
The amplification of magnetic seed fields seems crucial in the build-up of galactic magnetic fields.
There can be several contributors to the enhancement of the magnetic field:
First, compression of the gas can lead to an increase in the magnetic field proportial to $\rho^{\alpha}$.
The case of isotropic compression would correspond to $\alpha=2/3$.
Secondly, random and turbulent motions can drive a small-scale dynamo within the halo, which leads to an exponential growth of the magnetic field by $e^{\gamma{}t}$.
In the simulations with higher resolution, smaller scales and gas motions are resolved, leading to a faster amplification of the magnetic field.
During the assembly of this galactic halo, the magnetic field is amplified on a timescale $1/\gamma$ of the order of $10^{7}$ yrs \citep[see][]{beck12}.
We refer to the simulations of \cite{beck12} for more details about the amplification process.

\begin{figure}
\begin{center}
  \includegraphics[angle=90,width=0.475\textwidth]{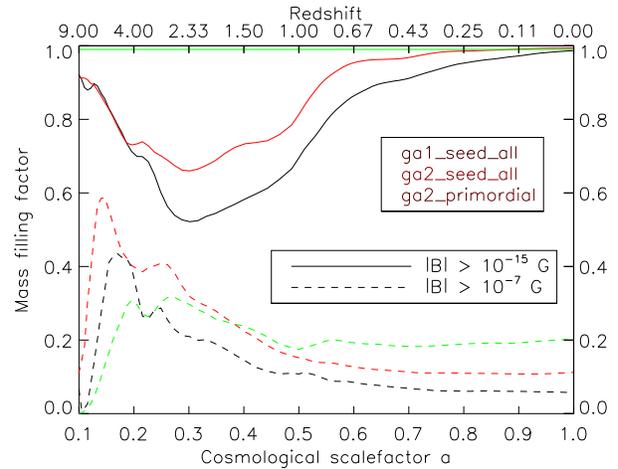}
  \caption{Mass fraction of the largest progenitor halo filled with magnetic fields.
The dashed lines correspond to strong magnetic fields and the solid lines to weak magnetic fields.
Most of the halo mass is filled with weak magnetic fields, however, strong magnetic fields are only contained within the mass at the halo center.}
  \label{fig:sim_mass}
\end{center}
\end{figure}

\begin{figure}
\begin{center}
  \includegraphics[angle=90,width=0.475\textwidth]{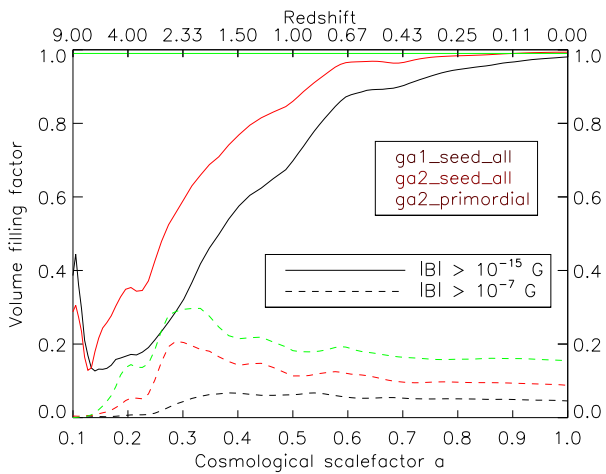}
  \caption{Volume fraction of the largest progenitor halo filled with magnetic fields.
The dashed lines correspond to strong magnetic fields and the solid lines to weak magnetic fields.
At first, only the innermost region of the halo where star formation takes place is magnetized.
Afterwards, diffusion and gas motions transport weak magnetic fields into the entire halo, however, strong magnetic fields are only contained within the innermost region at the halo center.}
  \label{fig:sim_volume}
\end{center}
\end{figure}

\begin{figure}
\begin{center}
  \includegraphics[angle=90,width=0.475\textwidth]{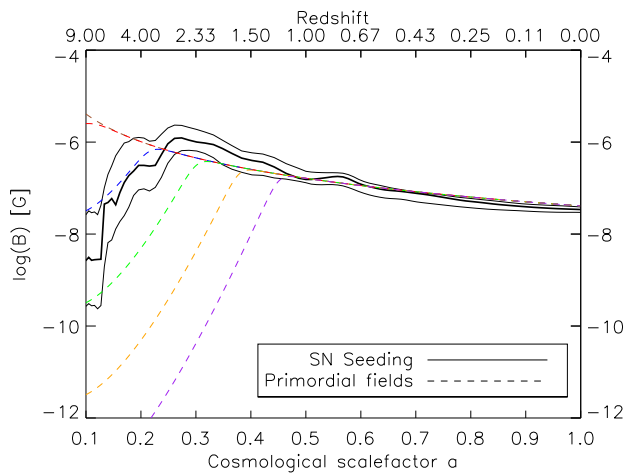}
  \caption{Various analytical model growth curves of \citet{beck12} adapted for our simulations.
The model curves represent the growth of the magnetic field amplitude within the diffuse halo gas for different initial primordial magnetic seed fields.
Overlaid (fat black lines) is the time evolution from our simulations with SN seeding for three different amplitudes.
The SN seed field is self-consistently created during the simulations.
The seed field is hence no longer a free parameter.}
  \label{fig:sim_model}
\end{center}
\end{figure}

\noindent{}Fig. \ref{fig:sim_mass} shows the fraction of magnetized halo mass against redshift in the simulation ga2\_seed\_all.
The size of the halo is determined by its virial radius and is dominated by the dark matter content.
At first, the halo is small in size and the majority of the gas content is located at the halo center and connected to starforming regions.
Within these starforming regions, the magnetic field is seeded and amplified, leading to a high magnetized mass fraction.
As the halo continues to grow in size, the mass magnetization fraction drops, presumably because the rate of accretion of unmagnetized gas is higher than the rate of magnetization.
Note that the gas mass content also changes as highly magnetized gas particles are converted into stars.
Until redshift z$\approx$0, diffusion and gas motions are able to remagnetize the entire galactic halo with magnetic fields stronger than $10^{-15}$ G.
However, magnetic fields stronger than $10^{-7}$ G are only contained within the innermost starforming region of the halo.
Within this densest region, seeding is still ongoing and dynamo action is most efficient here in maintaining strong magnetic fields.

\noindent{}Fig. \ref{fig:sim_volume} shows the fraction of magnetized halo volume against redshift in the simulation ga2\_seed\_all.
At first, most of the gas mass is located at the halo center, leading to a high magnetized gas fraction, but a low magnetized volume fraction.
Throughout the simulation seeding and amplification of the magnetic field up to a few $\mu$G occur only in the innermost region of the halo.
Thus, magnetic fields stronger than $10^{-7}$ G are only reached in the innermost parts of the halo volume.
However, the total magnetized halo volume increases significantly over time as the magnetic field diffuses towards the halo outskirts.

\noindent{}Independent of resolution, at redshift z$\approx$0 the entire halo mass and volume are magnetized.
However, in the simulation with higher resolution the magnetization occurs faster.
In Fig. \ref{fig:sim_amp} we showed that in the simulations with higher resolution the magnetic field is amplified faster and it also reaches higher saturation values.
Thus, the Alfv\'{e}n speed reaches higher values earlier in the simulations and also its maximum value is higher.
Because the diffusion speed is coupled to the sound- and Alfv\'{e}n speed (see section 2) we expect diffusion to behave differently.
Thus, it seems plausible that the magnetization of the halo mass and volume occur faster when increasing the numerical resolution.

\noindent{}We want to note, that our simulations do not include cosmic ray dynamics \citep[see e.g.][]{hanasz09} or an explicit model for galactic winds \citep[see e.g.][]{springel03}.
Cosmic ray driven winds can attain high velocities \citep[see e.g.][]{breitschwerdt91,reuter94,newman12} and could magnetize a galactic halo and the IGM quiet fast \citep{kronberg99} or even the largest voids \citep{beck13}.
The addition of an explicit model for galactic winds or a cosmic ray energy budget and pressure component into our simulations could cause additional gas motions.
Then the diffusion of the magnetic field would behave differently and could proceed much faster.
In the future, it will be worth to pursue cosmological simulations including cosmic rays and magnetic fields.

\noindent{}Fig. \ref{fig:sim_model} shows the redshift evolution of the halo magnetic field of our simulations with SN seeding and model growth curves of primordial fields.
The model growth curves reflect the time evolution of magnetic fields for different strengths of the primordial seed field, as found in simulations of the formation of a galactic halo \citep[for more details see][]{beck12}.
Various mechanisms exist for the generation of a primordial magnetic field, leading to a free and ambiguous parameter, with so far only very weak constraints from observations.

\noindent{}In our SN seeding model, no free parameter is left, as the seeding of the magnetic field is self-consistently coupled to the underlying starformation and the associated magnetic field strength seeded by each SN is taken from the observed value of magnetic fields in SN remnants.
The distribution and amplitude of the magnetic seed field therefore is the result of the underlying starformation process and the distribution and strength of starformation during structure formation.
We note that primordial seeding mechanisms may still operate and also lead to initial magnetic fields outside starforming regions or even create magnetic fields before the formation of stars.
However, as soon as the first magnetic fields are seeded by SN, the resulting magnetic fields within the associated structures by far exceed the contributions of primordial magnetic fields.

\subsection{High redshift rotation measures}

\begin{figure*}
\begin{center}
  \includegraphics[width=0.95\textwidth]{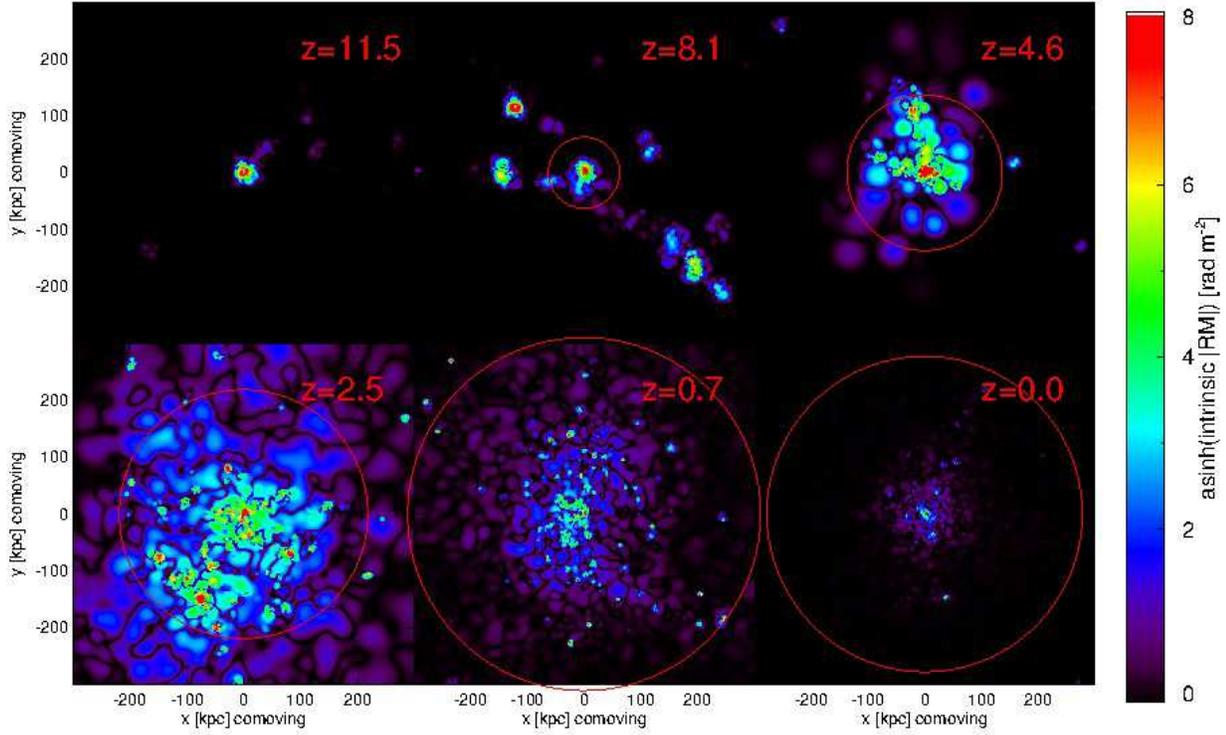}
  \caption{Intrinsic RM at different redshifts in the simulation ga2\_seed\_all of the forming galactic halo on an ``asinh'' scale for better visibility.
The contour plots can be interpreted in terms of a composition of Figs. \ref{fig:sim_dens} and \ref{fig:sim_bfld}.
Intrinsic RM values exceeding 1000 rad m$^{-2}$ can be found in starforming, strongly magnetized and dense gas clumps of the assembling halo at high redshift.
Towards redshift z$\approx$0 the halo virializes and the intrinsic RM values decline.}
  \label{fig:sim_fr}
\end{center}
\end{figure*}

\begin{figure}
\begin{center}
  \includegraphics[angle=90,width=0.475\textwidth]{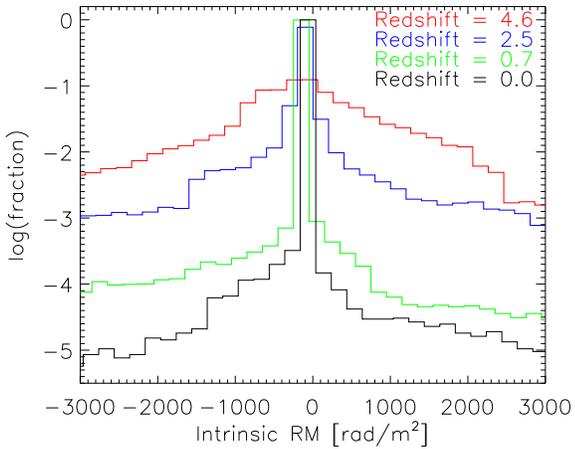}
  \caption{Histograms of the halo intrinsic RM in the simulation ga2\_seed\_all for different redshifts.
At high redshift, a large fraction of the halo hosts intrinsic RM exceeding 1000 rad m$^{-2}$, corresponding to a heterogeneous distribution of strongly magnetized and dense gas.
Towards redshift z$\approx$0 the halo virializes, the gas density and the magnetic field decline, substructures have disappeared and the intrinsic RM values decline (corresponding to Fig. \ref{fig:sim_fr}).}
  \label{fig:sim_histo}
\end{center}
\end{figure}

\begin{figure}
\begin{center}
  \includegraphics[angle=90,width=0.475\textwidth]{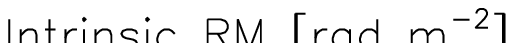}
  \caption{Evolution of the mean intrinsic RM in the simulation ga2\_seed\_all with redshift.
We show the mean intrinsic RM (smoothed with a gaussian beam of 10 kpc) of the innermost ten percent of the halo.
The colored region represents, on a percent-level spacing, the percentiles of the intrinsic RM distribution.
At redshift z$\approx$3 the RM reaches values of about 1000 rad m$^{-2}$.
Towards redshift z$\approx$0, the mean value declines significantly and becomes as low as 10 rad m$^{-2}$.}
  \label{fig:sim_RM}
\end{center}
\end{figure}

\noindent{}In this section we discuss the redshift evolution of the intrinsic RM of the forming galactic halo in our simulation.
The RM is calculated as the integrated line of sight product of the electron density $n_e$ and the magnetic field component $B_\parallel$ parallel to the line of sight $l$:

\begin{equation}\mathrm{RM}\sim{}\int{n_{e}B_\parallel}dl.\label{equ:RM}\end{equation}

\noindent{}Within the simulations the exact length of the integration path does not play an important role, as long as the halo is well inside the integrated line of sight.
We use a comoving line of sight of 200 kpc, which covers the virial radius of the forming galactic halo at all redshifts.

\noindent{}The evolution of the magnetic field distribution within the simulated galactic halo will also leave its imprints in the evolution of the RM.
Fig. \ref{fig:sim_fr} shows the maps of the intrinsic RM signal of the halo at six different redshifts in the simulation ga2\_seed\_all together with a circle, corresponding to the virial radius of the forming galactic halo.
In general, within the densest (see Fig. \ref{fig:sim_dens}) and strongest magnetized (see Fig. \ref{fig:sim_bfld}) regions we also find the largest intrinsic RM values.
Some of the values even exceed 1000 rad m$^{-2}$.
At the halo outskirts we find values of a few 10 rad m$^{-2}$.
Most interesting, the central protogalactic starforming region at redshift z$\approx$12 already hosts intrinsic RM values of several 1000 rad m$^{-2}$, indicating that strong magnetic fields and large RM must result during the formation of the first protogalaxies.
Fig. \ref{fig:sim_histo} shows the distribution of the intrinsic RM values within the halo for different redshifts.
At high redshift, the distribution of the intrinsic RM is widespread, corresponding to a heterogenous distribution of the gas within the halo during its assembly.
By virialization towards z$\approx$0 the magnetic field amplitude decreases and thus also the intrinsic RM values decrease.
The distributions at high and low redshift differ significantly.
We assume this to be caused by the process of halo formation and virialization.
At high redshift, the halo is in the process of formation and the distribution of dense and magnetized gas is highly heterogeneous, leaving its imprint in the intrinsic RM signal.
Towards redshift z$\approx$0, the halo virializes and the gas density as well as the magnetic field strength decline and also the intrinsic RM values decline.

\noindent{}Fig. \ref{fig:sim_RM} shows the evolution of the intrinsic RM value in the simulation ga2\_seed\_all with redshift.
We show the mean intrinsic RM value of the innermost ten percent of the largest progenitor halo.
The underlying colored regions mark the percentiles of the intrinsic RM distribution (percent-level spacing) and the outer-most contours mark the region of the 0.05-0.95 percentiles.
We smoothed our simulated synthetic intrinsic RM data with a beam size of 10 kpc to produce the mock data.
At redshift z$\approx$3, the RM reaches values of about 1000 rad m$^{-2}$.
Towards redshift z$\approx$0, the mean value declines significantly and becomes as low as 10 rad m$^{-2}$.

\noindent{}Given the extremely complex nature of RM observations, we do not intend to perform a comparison of our simulation with observations.
Furthermore, our simulation represents a single isolated evolving galactic halo, while the limited sample of observational data points at high redshift reflects a wide range of gas environments.


\section{Summary}

\noindent{}In this paper we present a model for the seeding and evolution of magnetic fields in protogalaxies. 
We introduce a numerical sub-grid model for the self-consistent seeding of galactic magnetic fields by SN explosions.
We perform cosmological simulations of Milky Way-like galactic halo formation including MHD, radiative cooling and star formation.
The main results are summarized as follows.

\noindent{}Within starforming regions, magnetic fields are seeded by SN explosions at a rate of about $10^{-9}$ G Gyr$^{-1}$.
In our simulations the first magnetic seed fields are created when the first stars form within the first protohaloes within the cosmic web.
Subsequently, the seeds are amplified by compression and turbulent dynamo action up to equipartition with the corresponding turbulent energy densities during the virialization of these first objects.
The random and turbulent motions are created by the gravitational collapse, SN activity and multi-merger events during the assembly of galactic haloes.
Within the hierarchical picture of structure formation, large objects form by mergers of multiple smaller objects.
As shown in simulations of idealized galactic mergers \citep[see][]{kotarba11,geng12a}, each merger event contributes to the amplification of the magnetic field.
This magnetic field is then also carried into the IGM by merger-driven gas motions and turbulent diffusion, providing a seed field outside starforming regions.
Furthermore, the IGM magnetic field is amplified by merger-induced shock waves, as well as possible dynamo action.
The final magnetic field strength reaches a few $\mu$G in the center of the halo and $\approx{}10^{-9}$ G at the halo outskirts (IGM).
The magnetic field distribution within the halo at redshift z$\approx$0 is comparable to the distribution obtained in simulations with primordial seeding \citep[see][]{beck12}.

\noindent{}The resulting magnetic field configuration is random and turbulent and additional galactic dynamo processes are necessary to produce a large-scale regular magnetic field topology.
In the assembly phase of the galaxy, the magnetic field can be amplified on timescales of a few ten million years \citep[see][]{beck12}.
In the case of primordial mechanisms the amplitudes of the magnetic seed fields are quite weak (order of $\approx{}10^{-18}$ G).
However, in the case of seeding by SN explosions, seed fields located within the SN remnants or -- more precisely -- within the resulting superbubbles of starforming regions are much stronger, with typical values of $\approx{}10^{-9}$ G \citep[see also][]{rees94}.
The presence of these stronger, non-primordial magnetic seed fields, lowers the amount of $e$-foldings required to reach $\mu$G amplitudes and hence shortens the time to reach equipartition magnetic fields within virialized haloes significantly.
However, the build-up of IGM magnetic fields is more challenging, as the magnetic field has first to be transported from the starforming regions into the IGM, before it can be subsequently amplified and further distributed.

\noindent{}At high redshifts, we find our simulated halo to host intrinsic RM values exceeding 1000 rad m$^{-2}$ within dense and highly magnetized regions.
The spatial distribution of those very large intrinsic RM is widespread, corresponding to the heterogenous distribution of starforming, and thus magnetized, gas within the halo during its assembly.
While the halo virializes towards redshift z$\approx$0 the gas distribution becomes more homogenous and also the halo magnetic field declines.
We find the intrinsic RM of our simulated halo to drop to a mean value below 10 rad m$^{-2}$ at redshift z$\approx$0.

\noindent{}Up to now, models for the evolution of cosmic magnetic fields always faced the problem of having the magnetic seed field as a free and ambiguous input parameter.
However, with our SN seeding model, the initial magnetic seed field is no longer a free parameter, but it is self-consistently created and described by the star formation process during the formation of cosmic structures.
Furthermore, the mean magnetic field generated by this mechanism in the protogalaxy by far exceeds the contribution of reasonable, primordial magnetic fields showing that primordial seeding mechanisms are not important in the context of galaxy formation.
Thus, our presented seeding model provides a general solution to the seed field problem within the context of galactic and cosmic magnetism.
We note that additional magnetic seed fields can still also be provided by AGN or primordial mechanisms.

\noindent{}So far, we only cover the evolution of the magnetic field within a forming galactic halo, as our simulations do not yet form a galactic disc at the halo center.
In the future, we plan to focus on the cosmological formation of disc galaxies within our simulations and study the detailed magnetic field structure within the evolving discs.

\noindent{}We conclude that the seeding of magnetic fields by SN and subsequent amplification during structure formation are able to build up strong magnetic fields of $\mu$G strength within very short timespans.
This leads to very strong magnetic fields within the very first collapsing and starforming protohalos at very high redshifts, which are the building blocks for the very first galaxies and could explain the observed, strong magnetic fields in galactic halos at high redshifts.
However, given the complex nature of star formation and MHD transport processes, we are still far from understanding the full spectrum of consequences and especially the imprint strong magnetic fields might impose on structure formation.


\section*{Acknowledgments}
We thank the anonymous referee for the report, which helped to improve the presentation of the paper.
We acknowledge additional comments from Uli Klein, Michal Hanasz, Rainer Beck and the members of the DFG Research Unit 1254.
We thank Stefan Heigl for proofreading of the paper.
Special thanks to Felix Stoehr for providing the original initial conditions.
KD is supported by the DFG Cluster of Excellence 'Origin and Structure of the Universe'.
PPK acknowledges support from an NSERC (Canada) Discovery Grant A5713.


\bsp

\label{lastpage}

\end{document}